\documentclass[article,twocolumn,amsmath,floats,showpacs,nofootinbib]{revtex4}
\usepackage[usenames]{color}

\usepackage{graphicx}
\usepackage{epstopdf}
\usepackage{textcomp}
\usepackage{hyperref}
\usepackage{multirow}

\hypersetup{colorlinks=true}

\begin{document}

\title{Reversible recovery of excess current in $Y(Ho)-Ba-Cu-O$-normal metal point contacts under high voltages}

\author{L.F. Rybal'chenko, V.V. Fisun, N.L. Bobrov, I.K. Yanson, A.V. Bondarenko, and M.A. Obolenskii}
\affiliation{B.I.~Verkin Institute for Low Temperature Physics and
Engineering, of the National Academy of Sciences
of Ukraine, prospekt Lenina, 47, Kharkov 61103, Ukraine and A. M. Gor'kov State University, Khar'kov
Email address: bobrov@ilt.kharkov.ua}
\published {(\href{http://fntr.ilt.kharkov.ua/fnt/pdf/17/17-2/f17-0202r.pdf}{Fiz. Nizk. Temp.}, \textbf{17}, 202 (1991)); (Sov. J. Low Temp. Phys., \textbf{17}, 105(1991)}
\date{\today}

\begin{abstract}A polarity-dependent reversible change in the current-voltage characteristics between states corresponding to different values of the excess current ${{I}_{exc}}$  is observed for bias voltages of several hundred millivolts in $Ag-ReB{{a}_{2}}C{{u}_{3}}{{O}_{7-x}}$ $\left( Re\text{ }=\text{ }Y,\text{ }Ho \right)$  point contacts in the current-carrying state. The observed effect has a high intensity (right up to a complete suppression or recovery of ${{I}_{exc}}$) and is not accompanied by a noticeable variation of the energy gap or the critical temperature in the point contact region. It is also established that the states with intermediate values of $I_{exc}$ are stable for zero bias voltages across the point contacts, at least up to $110 \ K$. It is assumed that the variation of ${{I}_{exc}}$ and the corresponding modification of IVC are due to the oxygen migration processes induced by the electric field and current, resulting in a local variation of the superconducting parameters in the point-contact region.

\end{abstract}

\maketitle

The complex structure of metaloxide high-temperature superconductors (HTS) predetermines their rather low stability to the action of various external factors of chemical, thermal, electromagnetic, or some other origin. In microscopically small point contacts (PC) with a direct conductivity, it is possible to obtain states characterized by high electric fields and current densities \cite{Yanson1}. Hence in experiments involving PC formed by HTS materials, the possibility of manifestation of electron migration processes in such materials cannot be ruled out. In the first place, one should expect variations of PC characteristics that are sensitive to the migration of oxygen in the crystal lattice of metaloxides, since the role of oxygen in the resulting superconductivity is vital in this case. Such a possibility was confirmed in the experiments by Ralls, Ralph, and Buhrman \cite{Ralls} on the observation of electron migration of defects in submicron copper bridges.

In the present work, we report on the observation of polarity-dependent reversible processes of switching over of current-voltage characteristics (IVC) between states with different excess currents in $\left( Ho \right)BaCuO-Ag$  point contacts for large voltages. The high intensity of the effect, which is manifested among other things in the attainment of states with a completely suppressed excess current and the accompanying variation of the PC conductivity from metallic to semiconductor type lead to the assumption that this effect may be due to the electron migration of oxygen in the crystal lattice of the HTS electrode.

The electrical characteristics (IVC and their first derivatives) of $Ag-ReB{{a}_{2}}C{{u}_{3}}{{O}_{7-x}}$ ($Re=Y$ and $Ho$) type point contacts created by a displacement technique \cite{Yanson1} on freshly cleaved surfaces of HTS were investigated. $YBaCuO$ samples were presented in the form of ceramics as well as single crystals, while $HoBaCuO$ samples were only in the form of single crystals. The samples were prepared by using the standard technique. The construction of mechanical devices used for fixing and mutual displacement of the electrodes made it possible to create new working points repeatedly on both electrodes without extracting them from the cryostat.

Measurements of the first IVC derivatives (differential resistance $dV/dI(V)$) were carried out by the standard modulation technique using the bridge method. Temperatures higher than $4.2\ K$ were obtained with the help of an intermediate cryostat connected with liquid helium through a capillary with a low flow rate and equipped with a low-voltage heater. This ensured the attainment of stable temperatures much higher than $100\ K$ for a low consumption of liquid helium. A superconducting solenoid immersed in liquid helium was used to produce magnetic fields with a maximum strength up to $\sim 50\text{ }kOe$. The mutual orientation of the PC and solenoid axes was not fixed clearly in our experiments, but the orthogonal configuration was maintained to within $20{}^\circ$.

Point contacts of $S-c-N$ type (\emph{c} stands for constriction) with a direct conductivity differ from tunnel junctions \cite{Artemenko,Blonder} in that their IVC display an excess
current $I_{exc}$ connected with the Andreev reflection of electrons (holes) at the $N-S$ boundary. This served as the main criterion in our selection of point
 contacts for experimental studies. The resistance of most of the contacts displaying the effect described in this paper was found to be within the interval
  $R_{N}=5-100~\Omega$ ($R_{N}$ is the differential resistance of the contact for $eV\gg \Delta $). In the dirty (Maxwell) regime of current flow through the contact \cite{Yanson1}, this interval will correspond to a contact diameter $d\simeq 1000-50\text{ }\AA$  for an HTS resistivity
    $\rho \text{ }\simeq {{10}^{-4}}\text{ }\Omega \cdot cm$. For pure contacts (Sharvin limit), the values of $d$ will be much lower although it is not possible
    to make any correct estimates in the absence of data on the constant  for $\rho_{l}$ HTS ($l$ is the electron mean free path).
    \begin{figure}[]
\includegraphics[width=8cm,angle=0]{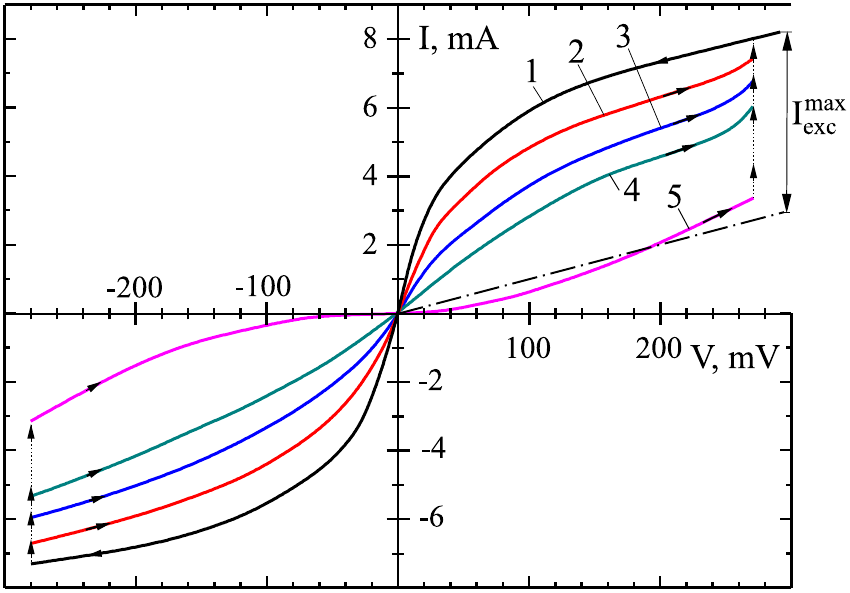}
\caption[]{Change in the IVC for states with different $I_{exc}$ in the case of a point contact of the type $HoBaCuO-Ag$ after the attainment of critical bias voltage $\left| {{V}_{cr}} \right|$ (the latter corresponds to the vertical dot-and-dash lines). The dot-and-dash straight line corresponds to the IVC expected for the $N$-state, $T=4.2~K$, $H=0$.}
\label{Fig1}
\end{figure}

 A typical set of IVC with different values of $I_{exc}$ for the same contact between $Ag$ and a single crystal of $HoBaCuO$ is presented in Fig \ref{Fig1} ($T=4.2~K$, $H=0$). The initial curve 1 corresponding to the maximum $I_{exc}$ is stable for both polarities of the bias voltages not exceeding $\sim 250\text{ }mV$. However, the situation changes radically for higher bias voltages across the contact. For example, as the bias voltage attains its critical values ${{V}_{cr}}\simeq -280\text{ }mV$  across the contact, a transition is observed to one of the other branches (curves 2, 3, 4, 5) with a lower, or even negative, excess current, depending on the time, for which the contact is subjected to this bias voltage. (Note that the sign of the bias voltage is opposite to that of the polarity of the HTS electrode). The transition rate also depends strongly on the bias voltage applied across the contact, and instant switching may be observed for quite large values of the bias voltage.

The reversal of the sign of bias voltage across the contact and its increase to values of the order of $\left| {{V}_{cr}} \right|$ results in a restoration of $I_{exc}$, i.e., a transition from curve 5 in Fig \ref{Fig1} to curves 4, 3, 2 or 1. In turn, each curve with an intermediate (curves 2, 3, 4) or completely suppressed $I_{exc}$ (curve 5) remains stable in time in the same range of bias voltage ($\sim 250~mV$) as the original curve, at least for a few hours. It should be emphasized that the transition between extreme states of the PC, i.e., between the states with maximum and zero values of $I_{exc}$ (curves 1 and 5 respectively in Fig. \ref{Fig1}) may occur with an indefinitely small level of discreetness which is determined only by the delay time and the proximity to which $V_{cr}$ is approached.

An analysis of unstable IVC for a large number (over 20) of PC based on $HoBaCuO$ single crystals with a wide range of resistance $R_N$ (from several ohms to tens of ohms) showed that it does not correlate with the value of $V_{cr}$. For the type of contacts considered by us, $V$ varied from 250 to 320 $mV$. Thus, the main factor determining a transition of the PC to states with different $I_{exc}$ is the voltage across the contact.

Note that the application of a magnetic field up to $\sim 45\ kOe$ had practically no effect on the value of $V_{cr}$ in spite of the fact that in many cases significant changes can take place in both $R_N$ and $R_0$ under the action of the field ($R_0$ is the differential contact resistance near $V$=0), i.e., in the entire IVC, at times even nonmonotonically. Note, however, that even in the strongest fields, the IVC of many contacts remained practically unchanged.

For $Ag-YBaCuO$ type point contacts, the main features of the effect described above are identical except for two cases. Firstly, the critical voltage $V_{cr}$ is quite high in the case of a ceramic as compared to that for a single crystal, and varies considerably from contact to contact. For some contacts, $V_{cr}$  was as high as 800 $mV$. Among other things, this is probably the reason why the effect considered by us could not be observed in earlier PC studies of the $YBaCuO$ ceramic \cite{Yanson2}. In the case of single crystals, the value of $V_{cr}$ was found to be nearly the same for both materials.

The second difference lies in the $YBaCuO$-based point contacts (both ceramics and single crystals) did not reveal any significant changes in their PC characteristics under the action of a magnetic field except in the low-energy interval ($eV<\Delta$) where various quantum coherence effects may emerge \cite{Yanson2}. Note, however, that not many experiments were carried out on single crystals of yttrium cuprate in a magnetic field.

However, it can be concluded from these experiments that the observed effect is not associated with the magnetic moments of any ions in the crystal lattice of yttrium or holmium cuprate.

\begin{figure}[]
\includegraphics[width=8cm,angle=0]{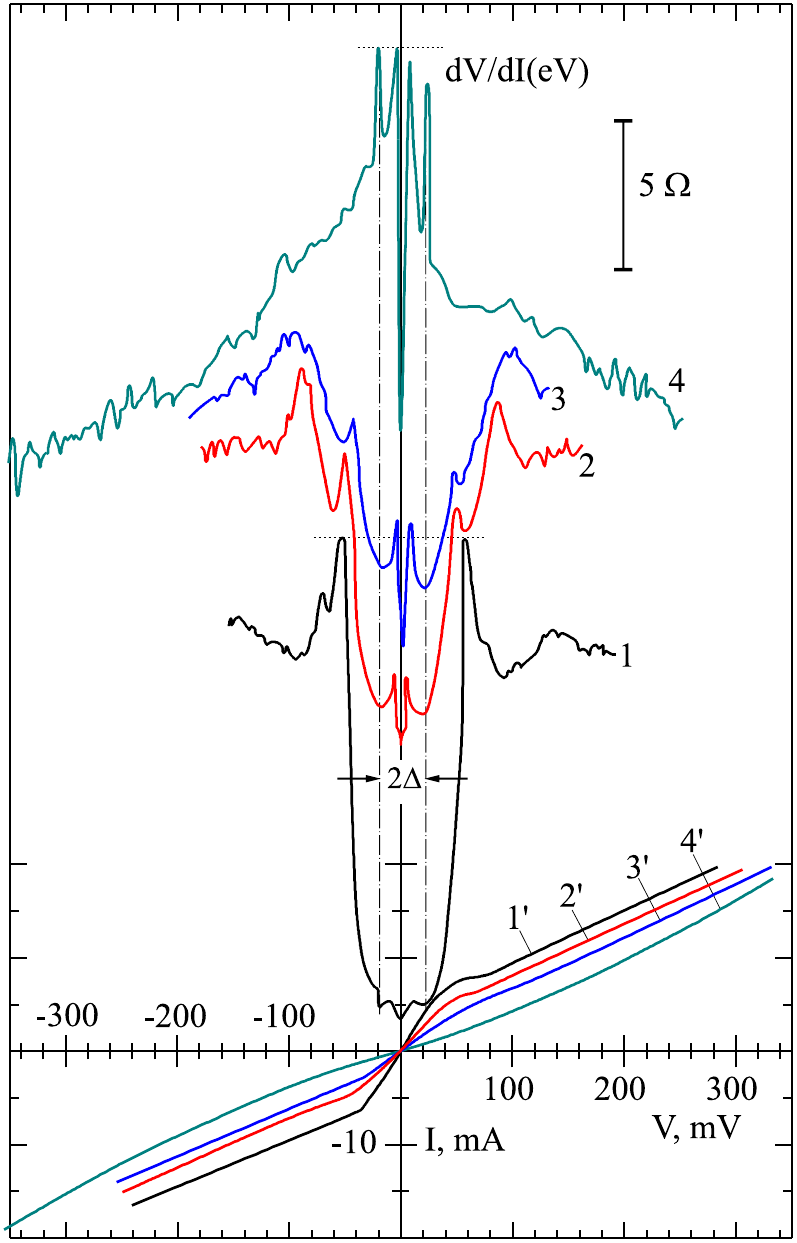}
\caption[]{Evolution of $dV/dI(V)$ dependences (curves 1-3) for a contact formed by a $HoBaCuO$ single crystal and $Ag$ with a decrease in $I_{exc}$, occurring at voltages close to - $V_{cr}$ (curves ${1}'-{3}'$  are the corresponding IVC); $T= 4.2~K, H=0$.}
\label{Fig2}
\end{figure}

The energy gap in $S-c-N$ contacts can be determined \cite{Artemenko,Blonder,Yanson3} quite easily from the positions of differential resistance minima on the $eV$ axis in the region of increasing $I_{exc}$. This circumstance makes it possible to establish a connection between $\Delta$ and $I_{exc}$ in states with a suppressed value of the latter.
Figure \ref{Fig2}
shows some intermediate IVC and  $dV/dI(V)$ for a typical contact based on $HoBaCuO$, which displays the reversible effect of suppression-restoration of $I_{exc}$. An analysis of these dependences leads to the important conclusion that no notable shift is observed in the position of gap minima along the voltage axis right up to the maximum suppression of $I_{exc}$. On the whole, the shape of the $dV/dI(V)$ dependence undergoes considerable variations. The most significant among these is an increase in the intensity of the zeroth anomaly (the differential resistance minimum near $V=0$) which accompanies a decrease in $I_{exc}$ and indicates \cite{Yanson3} the existence of weak bonds in the HTS electrode near the PC constriction. Moreover, with decreasing $I_{exc}$, the PC conductivity in the region $eV>\Delta$ becomes clearly of the semiconductor type.

\begin{figure}[]
\includegraphics[width=8cm,angle=0]{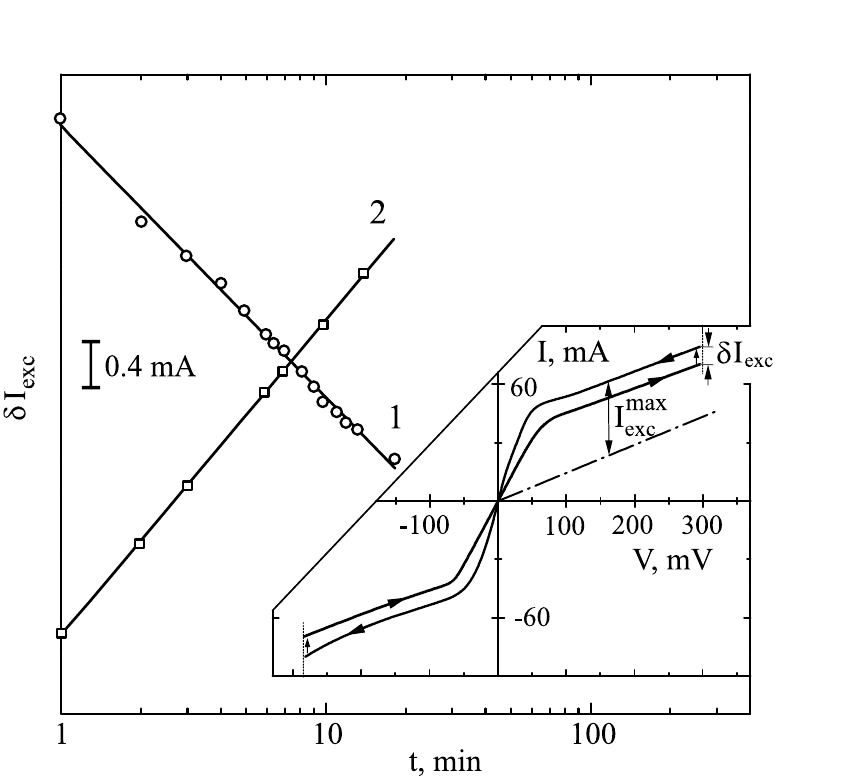}
\caption[]{Time dependences of the excess current variation $\delta I_{exc}$ near critical bias voltages $\pm300~mV$ for a $HoBaCuO$ single crystal-$Ag$ point contact: 1) suppression of $I_{exc}$ for $V=-300~mV$; 2) restoration of $I_{exc}$ for $V=+300~mV$. The inset shows the IVC corresponding to the initial value of the excess current; $T=4.2~K$, $H=0$.}
\label{Fig3}
\end{figure}

The dynamics of the $I_{exc}$ relaxation processes for a constant bias voltage in the instability region, i.e., in the vicinity of $V_{exc}$, was studied by recording the time dependence of the suppression and restoration of $I_{exc}$  in this region. The specific value of the bias voltage $V_{b}<V_{cr}$ was chosen so that the duration of the relaxation process was 5-20 min. For smaller durations, the process could become nonmonotonic, while larger durations were found to be inconvenient from the experimental point of view. An analysis of the experimental time dependences of the growth (decrease) of excess current for relatively small variations ($10-30\%$) near its peak value (see inset to Fig. \ref{Fig3}) showed that they follow quite precisely (Fig. \ref{Fig3}) the logarithmic dependence $\delta I\sim \ln t$ which is similar to the magnetic moment relaxation $\delta M\sim \ln t$ near $T_c$ due to a nonideal Meissner screening in a HTS (see, for example, Ref. \cite{Fiorani}).

Measurements of IVC and $dV/dI(V)$ dependences for contacts with partially suppressed $I_{exc}$ in a wide temperature interval showed that the states are stable at least up to 120 $K$ and the value of $I_{exc}$ remains unchanged upon a return to the initial temperature if the voltage applied across the contact is kept below the value $V_{cr}$. It was found that for states with partially suppressed $I_{exc}$, the value of $T_c$ determined for the contact region of the HTS electrode from the first derivatives of the IVC does not undergo any significant changes as compared with the unperturbed state. The observed decrease does not exceed 1-2 $K$.

\begin{figure}[]
\includegraphics[width=8cm,angle=0]{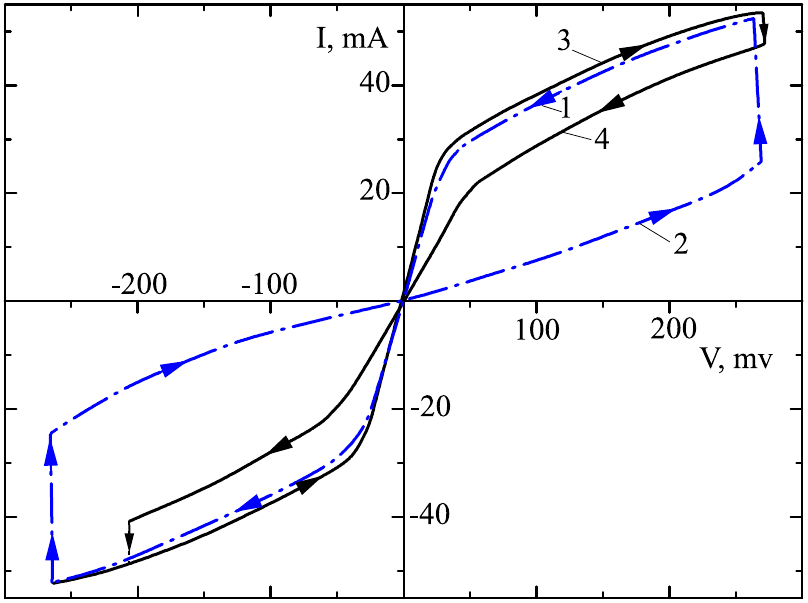}
\caption[]{Simultaneous observation of both (direct and inverse) transition effects between IVC with different $I_{exc}$ upon the attainment of critical bias voltage $|V_{cr}|$ across the $YBaCuO$ single crystal-$Ag$ point contact at $T=4.2~K$, $H=0$. The dot-and-dash lines (curves 1 and 2) show the behavior of the IVC for the direct effect, and the solid lines (curves 3 and 4) show the same for the reverse effect.}
\label{Fig4}
\end{figure}

Apart from the effect described above (we shall call it the direct effect), we also observed the opposite effect which is reversible like the direct effect. In the reverse effect, a voltage of sign opposite to that for the direct effect is applied across the contact for the suppression-restoration processes of $I_{exc}$. In Fig. \ref{Fig4},
the reverse effect is described by the IVC shown by solid lines, while the dashed lines correspond to the direct effect (in both cases, the same contact was used from the series $YBaCuO$ single crystal-$Ag$). It should be noted that there is a considerable difference in the critical voltages corresponding to the variation in $I_{exc}$ for both effects on the left half-plane while $V_{cr}$ for the reverse effect on the right half-plane is nearly the same or slightly exceeds the voltage corresponding to the direct effect. This circumstance made it possible to successively observe both effects on the same contacts in many cases. For a positive polarity of the $Ag$ electrode (right half-plane in Fig. \ref{Fig4}), the transition to the reverse effect corresponding to the suppression of $I_{exc}$ could begin only after the restoration of $I_{exc}$ was over in the direct effect. Note that a considerable difference in the values of $V_{cr}$ on the negative axis of bias voltages for the direct and reverse effects (Fig. \ref{Fig4}) is rather an exception than a rule. Apparently, this hampers not only a simultaneous emergence of both effects, but also of the reverse effect along in most contacts. Moreover, a significant feature of the reciprocal effect was the impossibility of a considerable variation of $I_{exc}$ in most of the contacts for which this effect was observed.

Let us consider the possible reasons behind the transition of IVC between states with different $I_{exc}$ for large bias voltages across the contact. It was mentioned above that the absence of a definite correlation between these effects and the external magnetic field rules out the quest for their magnetic origin. Considering the similarity of relaxation of $I_{exc}$ in point contacts and of magnetic moments in spin glasses, the concept of superconducting glass could be used as a model of HTS in which the variation of $I_{exc}$ could be attributed to the action of an electric field (current) on the phase synchronization of the order parameter in different weakly coupled crystallites in the HTS electrode. However, the fact that states with a suppressed $I_{exc}$ remain stable even after heating the contact to temperatures above $T_c$ does not lend credence to this concept also.

In an earlier work \cite{Rybal'chenko}, we reported on the transition from a state with maximum $I_{exc}$ to a state with minimum $I_{exc}$ in a $BiSrCaCuO$ based point contact under high voltages (several hundred $mV$) of either sign. This effect, which depends only on the magnitude (and not the sign) of bias voltage across the contact, also has some other distinguishing features. Firstly, no states with intermediate values of $I_{exc}$ are involved and, secondly, relaxation to the original state can occur either as a result of a partial decrease or a complete removal of the contact voltage without any change in its sign. This effect, which is essentially of hysteresis type, can be associated with the Josephson nature of the bonds between $Cu-O$ planes in $BiSrCaCuO$, so that the structure realized in the PC constriction is a chain of series-connected tunnel-type Josephson junctions. However, it is obvious that such an approach cannot be used to describe the results obtained in the present work.

We assume that the relaxation effects in \emph{Y(Ho)-Ba-Cu-O} point contacts observed in our investigations are associated with the oxygen subsystem, including the migration of oxygen in the crystal lattice of an HTS stimulated by an electric field and current. Leaving aside the reverse effect for the present, the suppression or restoration of $I_{exc}$ can be associated with a decrease or an increase, respectively, in the oxygen concentration in the PC constriction region from the side of the HTS electrode. The transition to the superconducting type of conductivity, which occurs simultaneously with a decrease in $I_{exc}$, can also be explained by this model. Since, as has been mentioned above, the suppression of $I_{exc}$ is not accompanied by any notable decrease in $\Delta$ or $T_c$, it can be assumed that the decrease in the oxygen concentration in the bulk of the PC constriction is local and not uniform.

Comparing the sign of polarity of the HTS electrode with the nature of variation of $I_{exc}$ for the direct effect, we can assume that the migration of oxygen occurs between the contact region of the HTS electrode and its inner layers. The increase in the intensity of the zeroth anomaly with decreasing $I_{exc}$ (Fig. \ref{Fig2}) is also an indirect proof of the same. As a matter of fact, the departure of oxygen from the PC region of the HTS electrode may also be accompanied by a weakening of the binding force between the crystallites, while it was shown earlier \cite{Yanson3} that the zero-order anomaly is an evidence of the existence of Josephson-type weak bonds in the PC constriction.

The high mobility of oxygen in metaloxides, for example, in $YBaCuO$, is also indicated by the data of Ref. \cite{List} in which photoemission measurements showed a decrease in the oxygen concentration at the cleavage surface even at 20 $K$. Moreover, according to some well-known works (see, for example, Ref. \cite{Ageev}), the thickness of the oxide film formed during the oxidation of metals is proportional to the logarithm of the reaction time. Since an analogous relation was obtained by us for the time dependence $\Delta I_{exc}(t)$, this coincidence can be treated as a further evidence in favor of our oxygen-migration model.

To explain the reverse effect, i.e., a partial decrease in $I_{exc}$ for a positive bias voltage across the HTS electrode, it should be borne in mind that the preliminary attainment of the state with maximum value of $I_{exc}$ is a necessary condition for this effect. A possible reason for this effect may be the increase in the oxygen concentration in the HTS lattice over the optimal stoichiometric value, caused by the migration of electrons. According to Ref. \cite{Song}, this leads to a structural disorder and, consequently, to a decrease in the critical parameters.

The absence of a correlation between the contact resistance and the critical voltage corresponding to the change in the excess current cannot be explained by the dirty (Maxwellian) model of the point contact, but an explanation is possible in the clean limit for which the current density is proportional to the voltage across the contact and is independent of its size (i.e., independence of $R_N$) \cite{Yanson1}. This is in agreement with the conclusions of Ref. \cite{Ralls} in which it was stressed that the electric migration of defects in point contacts is stimulated to a considerable extent by the processes of their collisions with electrons, i.e., depends directly on the current density. For such an approach, the electric field cannot play a dominating role in initiating this effect. Moreover, it becomes clear that the values of $V_{cr}$ observed for ceramic-based PC must be higher. Indeed, such contacts are unlikely to be clean. Consequently, for identical values of resistance, their size is larger than of pure contacts, and hence the critical current density corresponding to the effect of variation of $I_{exc}$ will be attained at higher voltages, as is indeed confirmed in the experiments.

Thus, for voltages of the order of a few tenths of a volt across $Y(Ho)-Ba-Cu-O$ point contacts, we have observed in the present work quite unusual processes of suppression of the excess current on the IVC for one polarity, and the restoration of its initial value for the opposite polarity. At least up to a temperature of $120~K$, the intermediate states involved in the excess current suppression are found to be thermally stable to a high degree. The emergence of such metastable states is not accompanied by a notable variation of the energy gap or the superconducting transition temperature in the contact region of the HTS electrode. The observed effects are attributed to a migration of oxygen in the crystal lattice of the HTS stimulated by an electric field and current.

\end{document}